\documentstyle[aps,twocolumn,psfig]{revtex}
\begin{document}

\draft
\tightenlines
 
\title{Avalanche and spreading exponents in systems with
absorbing states}
\author{
Miguel A. Mu\~noz$^{1,2}$,
Ronald Dickman$^{3}$,
Alessandro Vespignani$^{1}$,
 and 
Stefano Zapperi$^{4}$}
\address{
$^1$ The Abdus Salam International Centre for Theoretical Physics (ICTP) 
P.O. Box 586, 34100 Trieste, Italy\\
$^2$ 
Dipartimento di Fisica e unit\`a INFM,
Universit\'a di Roma `` La Sapienza", Piazzale A. Moro 2,
I-00185 Roma, Italy\\
$^3$ Departamento de F\'{\i}sica,
Universidade Federal de Santa Catarina,
Campus Universit\'ario\\
Trindade, CEP 88040-900,
Florian\'opolis --- SC, Brazil\\
$^4$PMMH-ESPCI, 
10, rue Vauquelin, 75231 Paris cedex 05, France \\
}
\date{\today}

\maketitle
\begin{abstract}
 We present generic scaling laws relating spreading critical exponents
and avalanche exponents (in the sense of self-organized criticality) 
in general systems with absorbing states.
Using these scaling laws we present a collection of
the state-of-the-art exponents for directed percolation, 
dynamical percolation and other universality classes.  
This collection of results should  help 
to elucidate the connections of self-organized criticality
and systems with absorbing states.
In particular, some non-universality
in avalanche exponents is predicted for systems with many 
absorbing states. 

\end{abstract}


\narrowtext

Directed percolation (DP) is broadly recognized as the
paradigmatic example of systems exhibiting a transition 
from an active to an absorbing phase \cite{Rev,Kin}.
DP critical behavior appears in
a vast array
of systems, among others
chemical reaction-diffusion models of catalysis \cite{catal},
the contact process \cite{Ha,Rev}, 
damage spreading transitions \cite{damage}, 
pinning of driven interfaces 
in random media \cite{inter},
roughening transitions in one-dimensional systems \cite{KW},
and Reggeon field theory \cite{RFT}.
This universality class has proven very robust with respect
to the introduction of microscopic changes, and many
apparently different systems share the same critical 
``epidemic" or ``spreading"  \cite{Torre} and ``bulk" exponents 
\cite{Rev,Kin}.
Nevertheless, examples
of a system exhibiting a transition to an absorbing state
outside the DP class  have been identified in recent years. 
Some examples are:
\begin{itemize}

\item Systems with 
two symmetric absorbing states or,  what is equivalent in 
many cases,
 systems in which the parity of the number of
particles is conserved \cite{Hin,BAW}.

\item Systems
with an infinite number of absorbing states, which exhibit 
nonuniversal spreading exponents \cite{PCP,INAS}.

\item Systems in which 
the dynamics is limited to the interface between active and
absorbing regions. These are in the class of the exactly solvable   
{\it voter model} \cite{liggett}, and compact directed percolation
\cite{CDP}.

\item Some models of epidemics
with immunization (no reinfection) \cite{DyP}. These  belong
to the so-called {\it dynamic percolation} class; the final set
of immune sites at criticality is a percolation cluster.

\end{itemize}
Recently, connections between
self-organized criticality (SOC) and systems with 
absorbing states have attracted much attention.
For example, there has been
a debate on whether the extremal Bak-Sneppen
for punctuated evolution  \cite{BS} and certain variants
are related to DP  \cite{BSDP}. It has also been argued
that sandpile models \cite{BTW}
share a number of features
with systems having many absorbing states \cite{noi},
and certain self-organized forest-fire models are related
to dynamical percolation \cite{Drossel}.

In self-organized models
the so called {\it avalanche exponents} are customarily
determined. 
Surprisingly, in spite of their obvious similarities,
the general connections
between spreading and avalanche exponents have not,
to the best of our knowledge, been given explicitly
for general systems with absorbing states.
Establishing the general scaling laws relating avalanche
and spreading exponents in systems with absorbing states
is the main goal of what follows.
This will allow us to put together many different
scaling relations and exponent values, presently quite
dispersed, and sometimes difficult to find in the literature,
and should facilitate progress in this field.

Let us first define in detail spreading and avalanche 
critical exponents.
The most accurate
determination of the critical point of systems with
absorbing states
comes from ``epidemic" or ``spreading" experiments
\cite{Torre}.
In these, a small perturbation (localized activity) is created at the
origin of an otherwise absorbing configuration, leading to a spread of
activity. In spreading experiments, it is customary to measure the
number of particles,
averaged over all runs (including those that have reached the
absorbing state) $ N(t)$,  the
survival probability $P(t)$,
and the mean-squared deviation from the origin
$R^2(t)$.
At criticality these magnitudes scale as
\begin{equation}
N(t)\sim t^{\eta};~~~~~~
P(t)\sim  t^{-\delta}; ~~~~~
R^2(t)\sim  t^z 
\label{se}
\end{equation}
where $\eta$, $\delta$ and $z$ \cite{zeta} are the spreading exponents. 
 
Once the critical point has been located accurately 
all the remaining standard critical indices can be estimated.
For reference we show in Table I 
a compilation of the state-of-the-art 
values for the usual exponents in directed percolation,
corresponding to a synthesis 
of the best series expansion and simulation results.
For the sake of completeness let us give here their respective
definitions. Calling $\Delta$ the distance to the
critical point in terms of the reduced control parameter,  
$\rho$ the order parameter, $\xi_{\perp}$
($\xi_{\parallel}$) the characteristic length (time), 
$h$ an external field coupled to $\rho$, 
and $\chi \equiv L^d {\rm var}(\rho)$ the static 
``susceptibility", 
we have: $\rho \sim \Delta^{\beta}$, 
$\rho  \sim h^{1/\delta_h}$ at the critical point,
$\chi \sim \Delta^{-\gamma'}$,
$\xi_{\perp}\sim \Delta^{-\nu_{\perp}}$,
$\xi_{\parallel}\sim \Delta^{-\nu_{\parallel}}$,
$P_{\infty} \equiv \lim_{t \to \infty} P(t) \sim \Delta^{\beta'}$,
and $\rho(t) \sim t^{-\theta}$ at the critical point.

>From the whole set of exponents that can be defined in DP,
only three are independent; the rest can be determined using 
well-known scaling relations (see appendix).
In certain systems possessing 
an infinite number of absorbing states \cite{PCP,INAS}, a fourth 
independent critical exponent has to be introduced \cite{JFFM,delta}.
This is due to the fact that the exponent $\delta$ (which in
DP coincides with $\theta$) is
non-universal and depends on the nature of the absorbing state 
in which the epidemic spreads \cite{JFFM,delta}.
(Similarly, the exponent $\beta'$, normally identical to $\beta$,
varies along with $\delta$ in such systems \cite{JFFM}.)

On the other hand, studies of avalanche transport employ a different
definition of the spread of activity.
For instance, in the prototypical sandpile model  \cite{BTW}
avalanches are obtained by adding one sand grain to a 
stable or {\it absorbing} configuration. In this way the system 
jumps among absorbing configurations via avalanche-like rearrangements.
The following quantities and associated 
exponents are usually measured:
\begin{eqnarray}
P(s)  & \sim &  s^{-\tau} f(s/s_c)  \\
s_c  &  \sim &  \epsilon^{-1/\sigma}    \\
\langle s \rangle & \sim &  \epsilon^{-\gamma} 
\end{eqnarray}
where $s$ is the size of an avalanche,
(total number of active or toppling sites),
$P(s)$ the associated probability distribution,
$s_c$ the cut-off size, 
$\langle s \rangle$ the mean size, and
$\epsilon$ represents the temperature-like variable associated with
the process: $\Delta$ for contact process or DP, the dissipation rate
for sandpiles, $F-F_c$ in driven-interface models (here $F$ is the
driving force). If $\epsilon=0$, the characteristic length is defined 
by the system size $L$  through the scaling relation $s_c\sim L^D$.
Analogously, the following exponents 
associated with the duration $t$ are also measured
\begin{eqnarray}
P(t)     & \sim &  t^{-\tau_t} g(t/t_c) \\
t_c      &  \sim &  \epsilon^{-1/\sigma_t} \\
\langle  t \rangle & \sim &  \epsilon^{-\gamma_t}.
\end{eqnarray}
Let us now  provide the general scaling laws relating 
avalanche and spreading exponents in systems with 
absorbing states.

>From the definitions of
$\eta$ and $\delta$, it is evident that
the total number of particles in {\it surviving} runs goes like
$N_s \sim t^{\eta + \delta}$, and therefore its time integral is governed
by the exponent $1+\eta + \delta $.
Thus an avalanche that dies at time $t$ has a typical size
$s \sim t^{1+\eta + \delta}$.
The probability to
die between times $t$ and $t+dt$ scales as $D(t) \sim t^{-\delta-1} dt$.
Observe that the time is defined in such a way that after
a `toppling' (updating of a given site), it is increased
by $\Delta t  = 1/N_s(t)$ \cite{updating}.
Therefore the number of topplings per unit time is $ N_s(t)$.
To express $\tau$, $\sigma$ and $\gamma$ as functions of the spreading 
exponents, let us consider a specific avalanche size, 
say $s_1$. An avalanche of size $s_1$ can have different 
durations, since $t$ and $s$ are not related in a deterministic way,
i.e.,
\begin{equation}
P(s_1) = \int_{t_1}^{t_2} dt P(s_1|t) D(t)  \;,
\end{equation}
where $t_1$ and $t_2$ are the minimum and maximum times compatible
with $s_1$, and $P(s|t)$ is the conditional probability of an avalanche
having size $s$, given it dies at time $t$.
$P(s|t)$ is bell-shaped, with its maximum at
$t \sim s^{1/(1 +\eta +\delta)}$\cite{notaste}.
Writing
$P(s|t) = t^{-(1 +\eta +\delta)} F(s/t^{1 +\eta +\delta})$,
where $F(u)$, the (normalized) scaling function, is nonsingular, we
have, on changing variables
\begin{equation}
P(s) = s^{-(1 +\eta + 2\delta)/(1 +\eta +\delta)}
\int du u^{\delta/(1 +\eta +\delta)} F(u)  \;,
\end{equation}
in other words,
\begin{equation}
\tau= {1+\eta+2\delta \over 1 +\eta +\delta}.
\end{equation}
With $t_c \sim \epsilon^{-\nu_{||}}$, and using
$s_c \sim t_c^{1 +\eta +\delta}$,
we have
\begin{equation}
1/\sigma= \nu_{\parallel} (1 +\eta +\delta).
\end{equation}

The remaining exponent $\gamma$ and the fractal dimension
$D$ can be determined using the
relations $\gamma=(2-\tau)/\sigma =\nu_{\parallel}
(1 +\eta)$ (this last equation for $\gamma$ has already 
been found by other authors \cite{IJ}),
and the standard relation $D=1/(\sigma \nu_{\perp})$.
Following a very similar derivation to the one 
just presented, one can easily determine also
the following scaling relations for the exponents associated
with $P(t)$:  
\begin{eqnarray}
 \tau_t & = & 1+\delta \nonumber \\
\sigma_t 
&=&\sigma (1+\eta +\delta)= 1/\nu_{\parallel} \nonumber \\
\gamma_t & = & (2-\tau_t)/ \sigma_t= \nu_{\parallel}
(1-\delta).
\end{eqnarray}

Let us derive explicitly the expression for $\gamma_t$.
In DP, we have, for $\Delta < 0$, or for $\Delta = 0$ and finite $L$,
the scaling form for the survival probability
$P(t) \sim t^{-\delta} e^{-t/t_c}$ with $t_c \sim |\Delta|^{-\nu_{||}}$,
or $t_c \sim L^{\nu_{||}/\nu_{\perp}}$.  Since the probability density
for dying at time $t$ is $- d P(t)/dt$, we can write
\begin{eqnarray}
\langle t \rangle  =  \int t \frac {d}{dt} [- P(t)] dt 
  &\sim & \int_{t_0}^{\infty} t^{-\delta} e^{-t/t_c} dt  \nonumber \\
  & \sim & t_c^{1-\delta} \int_{u_0}^{\infty} u^{-\delta} e^{-u} du.
\end{eqnarray}
So $\langle t \rangle \sim |\Delta|^{-(1-\delta)\nu_{||}} $, giving 
the value of $\gamma_t$. 
In the finite-size case,
$\langle t \rangle \sim L^{-(1-\delta)\nu_{||}/\nu_{\perp}} $
(observe that $t_0$ and $u_0$ are unimportant lower cut-offs).

{\it All the scaling relations derived so far are general,
and valid for all systems with absorbing states}.
Specific scaling relations for systems in the DP class
can be written using the well known relation \cite{Torre}
$\eta + 2 \delta  = d z/2$. 
Using the best values for the spreading exponents in DP 
taken from the bibliography (Table I), we obtain the values
of the avalanche exponents for DP 
in different dimensions (they are also summarized in table I).

Applying our general relations to other classes of models, 
we obtain:

\begin{itemize}

\item For models with parity conservation, using the known
result for spreading exponents \cite{BAW}, we predict
$\tau \approx 1.22$, $\sigma \approx 0.24$, $\gamma
\approx 3.25$, $\tau_t \approx 1.28$,
$\sigma_t \approx 0.31$ and $\gamma_t \approx 
2.33$ in $d=1$ \cite{ra} and mean field 
values above that dimension. These results have also been derived
and numerically tested in \cite{rc}. 

\item 
In systems with many absorbing states,
a generalized    
hyperscaling relation has to be introduced, due to the
fact that in this case the exponents $\delta$ and $\eta$
are non-universal and
therefore $\delta \neq \theta$ in general (on the other
hand the combination $\eta +\delta$ retains
its DP value). The generalized scaling 
law for these systems is \cite{JFFM,delta}
$\eta +\delta +\theta = d z/2$. 
Applying our scaling laws we predict  non-universal
values of $\tau$, $\gamma$, $\tau_t$ and $\gamma_t$
for systems with many absorbing states; i.e. if experiments are   
performed on a fixed environment, the results depend upon the 
environment itself. 
Recently this kind of non-universality has been 
observed in the class of sandpile models with fixed energy 
\cite{FES}. 
  
\item For models in the CDP class we have $\tau= 4/3$, 
$\sigma=2/3$, $\gamma=1$, $\tau_t=3/2$, $\sigma_t=1$
and $\gamma_t= 1/2$ in $d=1$  and mean field values in $d=2$ and
above.

\item For dynamical percolation
we can take advantage of our scaling laws, 
using them the other way around, i.e.,  
using the well known avalanche (cluster)
exponents for standard percolation \cite{ndyp,havlin}
permits us to determine 
the spreading exponents \cite{delta} with good accuracy. 
In table II, we report a collection of exponent values in 
$d=2,3$ and $6$ spatial dimensions. 
\end{itemize}

In summary, we have presented the general scaling relations 
that rule general systems with absorbing states, and present 
a collection of exponent values that can be useful as a reference. 
We believe that this coherent derivation and collection of otherwise 
scattered scaling laws and exponent values may
facilitate progress in drawing connections and similarities in 
many systems which show absorbing states and avalanche behavior.

\vspace{1.0cm}

{\it ACKNOWLEDGMENTS-}  This work has been partially 
supported by the European network, project number
FMRXCT980183. We thank K. B. Lauritsen, H. Park and R.M. Ziff 
for useful comments and remarks.

After completion of this work we became aware 
of a recent paper by Lauritsen et al. \cite{new}, in which 
very similar scaling relations to the ones proposed here are
derived for directed 
percolation in the presence of an absorbing
wall. In particular, as in systems with many absorbing
states they find in that case
$\delta \neq \theta$. A direct consequence
is that, as discussed here, some avalanche exponents do not
take their corresponding DP values. 
Some other interesting scaling relations can be found in \cite{Maslov}.

\vspace{0.5cm} 

{\bf APPENDIX: Scaling relations for DP} 

\vspace{0.2cm}

Here we present a collection of scaling laws for the DP 
universality class\cite{Rev,Kin,Torre}.
\begin{eqnarray}
\eta + \delta + \theta & = & d z /2    \nonumber \\
\delta_h  & = & (\nu_{||} + d\nu_{\perp})/\beta -1   \nonumber \\
\beta & = & \beta'  \nonumber \\
\delta & = & \theta  \nonumber \\
\gamma' & = & \gamma -\nu_{||} =  d \nu_{\perp} - 2 \beta
 \nonumber \\
\beta & = & \theta \nu_{||}  \nonumber \\
\beta' & = & \delta  \nu_{||}  \nonumber \\
z&=&2 \nu_{\perp} /  \nu_{||} \nonumber \\
D&=& 1/(\sigma \nu_{\perp})= d+(\nu_{||} -\beta)/\nu_{\perp}\nonumber \\
\gamma &= & (2-\tau)/\sigma = d \nu_{\perp} + \nu_{||} - 2 \beta. 
\end{eqnarray} 
Observe that not all of these relations are independent.

\newpage

\begin{table}
\begin{center}
\begin{tabular}{|c|l|l|l|l|}
\hline
$exponent$     & $d=1$          &    $ d=2 $    &    $ d=3 $     &
 $d=4$ \\
\hline
\hline
 $\beta = \beta'$     & 0.27649(4)$^a$ & 0.583(4)$^c$   &   0.805(10)$^g$ &  1
 \\
  $1/\delta_h$ &   0.111(3)$^b$ & 0.285(35)$^b$  &   0.45(2)$^h$ &  1/2
\\
 $\gamma'$    & 0.54386(7)$^a$ & 0.35           &   0.19      &  0   \\
 $\nu_{||}$  & 1.73383(3)$^a$ & 1.295(6)$^d$   &   1.105(5)$^g$  &  1
 \\
 $\nu_{\perp}$ & 1.09684(1)$^a$ & 0.733(4)$^e$   &   0.581(5)  &  1/2 \\
$\delta =\theta $  & 0.15947(3)$^a$ & 0.4505(10)$^f$ &   0.730(4)$^g$  &  1
 \\
 $ \eta   $  & 0.31368(4)$^a$ & 0.2295(10)$^f$ &   0.114(4)$^g$  &  0
 \\
 $ z      $  & 1.26523(3)$^a$ & 1.1325(10)$^f$ &   1.052(3)$^g$  &  1
 \\
$\nu_{||}/\nu_{\perp}$& 1.58074(4) & 1.766(2)       &   1.901(5)  &  2
\\
\hline
\hline
$\tau$     & 1.108  & 1.268 & 1.395 & 3/2 \\
$\sigma$   & 0.391  & 0.459 & 0.490 & 1/2 \\
$\gamma$   & 2.277  & 1.593 & 1.232 & 1 \\
$D_f $     & 2.328  & 2.968 & 3.507 & 4 \\
$\tau_t $  & 1.159  & 1.450 & 1.730 & 2 \\
$\sigma_t$ & 0.576  & 0.771 & 0.904 & 1 \\
$\gamma_t$ & 1.457  & 0.711 & 0.298 & 0 \\
\end{tabular}
\end{center}
\label{cpmom}
\noindent {\small Critical exponents for Directed Percolation. Exponents 
calculated by using scaling relations contained in this paper are reported 
in the lower part. $^a$ See \cite{IJ};
$^b$ \cite{AD}; 
$^c$ obtained using $\beta = \delta \nu_{||}$;
$^d$  \cite{GZ}; 
$^e$ obtained using $\nu_{\perp} = z \nu_{||}/2 $;
$^f$  \cite{VZ};
$^g$ \cite{IJ2};
$^h$   \cite{A2}. Where not reported uncertainties are in the last digit. 
For $d=4$ we report the exact mean field values.}
      
\end{table}
\begin{table}
\begin{center}
\begin{tabular}{|c|c|c|c|}
\hline
$exponent$     & $d=2$ &    $ d=3$    &    $ d=6 $ \\
\hline
\hline
 $\beta = \beta'$ & 5/36 & 0.417   & 1 \\
 $\nu_{||}$  & 1.506 & 1.169 &  1 \\
 $\gamma$   & 43/18  & 1.795 & 1 \\
 $\nu_{\perp}$ & 4/3& 0.875  & 1/2 \\
$\tau$ & 96/91  & 1.188 & 3/2 \\
$\sigma$   & 36/91  & 0.452 & 1/2\\
$D_f $     & 91/48  & 2.528 & 4 \\
\hline
\hline
$\tau_t $  & 1.092  & 1.356    & 2 \\
$\sigma_t$ & 0.664  & 0.855 & 1 \\
$\gamma_t$ & 1.367  & 0.752 & 0 \\
$\eta   $  & 0.586  & 0.536 &  0\\
$\delta =\theta $  & 0.092 & 0.356 &  1\\
$ z      $  & 1.771 & 1.497 &  1\\
\end{tabular}
\end{center}
\label{dymom}
\noindent {\small Critical  exponents for dynamical percolation. Exponents 
calculated by using scaling relations contained in this paper are reported 
in the lower part. The rest of  exponents values are from \cite{havlin}.
Where not reported uncertainties are in the last digit. For $d=2$, values
expressed as fractions refer to exact results\cite{havlin}. For $d=6$ we 
report the exact mean field values.}
      
\end{table}



\begin{thebibliography}{99}
\bibitem{zeta}  Observe that this is not the dynamic exponent $z_d$
defined usually as $t_c\sim\xi_{\perp}^{z_d}$. The two exponents are 
related by $z=2/z_d$.
\bibitem{updating} This is for sequential updating. 
For parallel processes time is increased in one unit after
each updating.
\bibitem{notaste}
We have verified numerically this assumption for DP as well 
as for a stochastic sandpile model.
When  this assumption is not satisfied, inconsistent results
can be obtained: A. Chessa, H.E.Stanley, A. Vespignani and S. Zapperi,
Phys. Rev. E {\bf 59}, R12 (1999).
\bibitem{ra} These exponents corresponds to the case in which
the absorbing state can be reached. If is not accessible, 
the exponent $\delta$ is zero \cite{BAW},
and the
exponents $\tau$, $\tau_t$, $\gamma$ and $\gamma_t$ are modified.
\end{thebibliography}
\end{document}